\documentclass[twocolumn,showpacs,preprintnumbers,amsmath,amssymb,pra]{revtex4}
  \usepackage{graphicx}
  \usepackage{dcolumn}
  \usepackage{bm}

\begin{document}


 \title{Measurement and simulation of laser-induced fluorescence from non-equilibrium ultracold neutral plasmas}

 \author{A. Denning$^1$, S. D. Bergeson$^1$, and F. Robicheaux$^2$}

 \affiliation{$^1$Department of Physics and Astronomy,
    Brigham Young University, Provo, UT 84602 \\
 $^2$Department of Physics, Auburn University, Auburn, Alabama 36849}

\date{\today}

\begin{abstract}
  We report new measurements and simulations of laser-induced fluorescence in ultracold neutral plasmas.  We focus on the earliest times, when the plasma equilibrium is evolving and before the plasma expands.  In the simulation, the ions interact via the Yukawa potential in a small cell with wrapped boundary conditions.  We solve the optical Bloch equation for each ion in the simulation as a function of time. Both the simulation and experiment show the initial Bloch vector rotation, disorder-induced heating, and coherent oscillation of the rms ion velocity.  Detailed modeling of the fluorescence signal makes it possible to use fluorescence spectroscopy to probe ion dynamics in ultracold and strongly coupled plasmas.
\end{abstract}


\maketitle

\section{Introduction}

Ultracold neutral plasmas can be formed by photo-ionizing
laser-cooled gases \cite{killian99, kulin00, simien04, cummings05a, fletcher06, killian07}.  The initial ion temperature is
approximately equal to the the laser-cooled atom temperature,
typically 0.01 K or less.  These plasmas are also small.  Their
characteristic size is typically 1 mm.  The electron temperature can be tuned from approximately 0.5 K to 1000 K.  At low initial
electron temperature, space-charge effects prevent electrons from
leaving, and the plasmas are charge neutral \cite{killian99}. The electrons screen the ion-ion interactions, playing an important role in the rate at which the ions thermalize the the plasma evolves \cite{killian01, kuzmin02a, kuzmin02b, mazevet02, robicheaux02, robicheaux03, bergeson03, roberts04, li04, pohl04a, pohl05c, cummings05b, pohl06, fletcher06, murillo06, laha07, gupta07, ivanenko07, bergeson08, rolston:2, zhang08a, zhang08b, castro09, zhang09}.

Immediately after photo-ionization, the ion-ion interaction raises
the ion temperature.  This happens because the ions are randomly
distributed throughout the plasma. Although they have essentially
zero kinetic energy, they have a great deal of electrical potential
energy.  The ions move to minimize this potential energy, a process called disorder-induced heating \cite{simien04, murillo06}.  The initial ion motion is coherent in a sense, and damped oscillations have been observed in the ion temperature evolution.

During this initial plasma phase, the ion temperature is not well
defined \cite{murillo06}.  This is somewhat surprising because the initial
nearest-neighbor distribution, which dominates the disorder induced
heating phase, is approximately Gaussian.  It has also been shown
that the spatial distribution evolves self-similarly (i.e.: is
always Gaussian if the ion temperature is not too low) \cite{robicheaux03}. Furthermore,
the velocity distributions both before and after the
disorder-induced heating phase \textit{are} Gaussian. Therefore one
might expect a nominally Gaussian distribution of velocities
throughout the thermalization time.

Early fluorescence and absorption studies of ions in these plasmas
used a Voigt profile to extract the ion temperature \cite{simien04,cummings05a}.  The Voigt
profile,
\begin{equation}
  S(\nu) = \int {L}(\nu-x^{\prime}) {G}(x^{\prime}) \; dx^{\prime},
  \label{eqn:voigt}
\end{equation}
is a convolution of the Lorentzian natural line shape $L(\nu)$ with
the Gaussian $G(\nu)$ thermal line shape.  While this appears to be
valid in the later stages of the plasma evolution, the Gaussian width
has limited meaning as the ions thermalize.  Recent work has shown that Eq. \ref{eqn:voigt} may be valid locally in the plasma \cite{castro09}.

In this paper, we present a study of laser-induced fluorescence in
ultracold neutral plasmas.  Rather than extract the temperature
using a lineshape analysis or local fluorescence imaging, we perform a simulation of the fluorescence by numerically integrating the optical Bloch equations
for the plasma ions.  The simulated fluorescence signal is compared
with experimental results.  Both the experiment and simulation show
the initial rotation of the Bloch vector as the ions (all initially
in the ground state) begin to absorb and scatter laser light.  During the
disorder-induced heating phase, the linewidth broadens and the
fluorescence signal falls.  At high densities, a coherent oscillation
in the ion motion is observed.  The experiment and
simulation have satisfactory numerical agreement, and we suggest
ways to use these computational techniques in future experiments in strongly-coupled neutral plasmas.

\section{Experiment}

Up to seven million $^{40}$Ca atoms are laser-cooled and trapped in a
magneto-optical trap (MOT).  The trap density is approximately
Gaussian of the form
\begin{equation}
  n(r) = n_0 \exp\left( -r^2/2\sigma^2 \right), \label{eqn:dens}
\end{equation}
where $n_0$ is the peak density with a typical value of $10^{10}$
cm$^{-3}$ in our experiment, and $\sigma$ is the one-dimensional
rms size of the MOT, typically 350 $\mu$m.

\begin{figure}[t]
  \includegraphics[angle=270,width=3.35in]{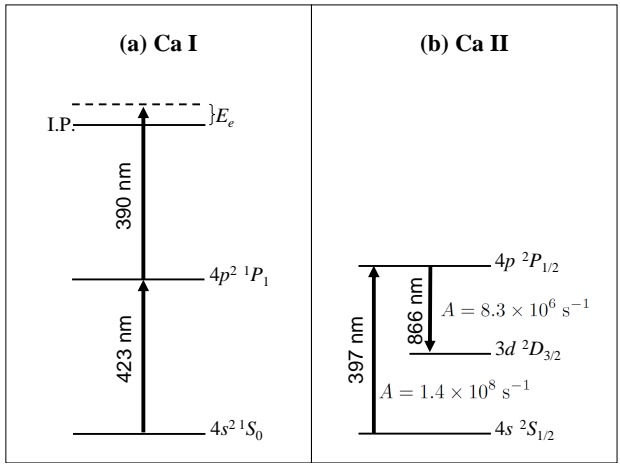}
  \caption{Partial energy level diagram, transition wavelengths, and transition probabilities for relevant levels in (a) Ca I and (b) Ca II.  The initial electron energy $E_e$ is determined by the energy difference between the photoionization laser photon energy and the Ca I ionization energy (I.P.).  The initial electron energy is $T_e = 2E_e / 3 k_B$.  The 423 nm Ca I transition is also used for the calcium MOT.  Not shown is a repumper laser at 672 nm that plugs an optical leak from the MOT.}
  \label{fig:level}
\end{figure}

Atoms in the trap are photoionized using ns-duration pulsed lasers
at 423 and 390 nm.  The temperature of the plasma ions is
approximately equal to the $\sim$mK temperature of the MOT atoms.
The initial energy of the plasma electrons is determined by the
wavelength of the 390 nm laser.  It is equal to the difference
between the laser photon energy and the atomic ionization potential
(see Fig. \ref{fig:level}).

\begin{figure}
  \includegraphics[width=3.35in]{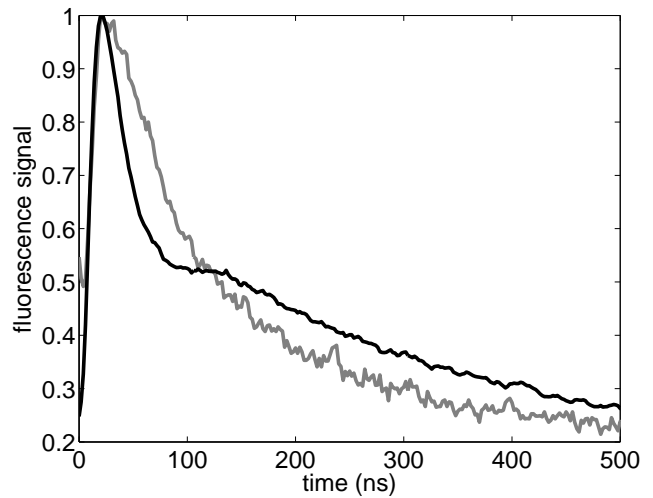}
  \caption{Relative fluorescence signals from a high density (black line, $n_0 = 7.3 \times 10^9 \; \mbox{cm}^{-3}$) and low density (gray line, $n_0 = 0.45 \times 10^9 \; \mbox{cm}^{-3}$) plasma at $T_e = 57$ K.  In the high density plasma, disorder induced heating occurs more rapidly.  The fluorescence signal shows a weak oscillation around 100 ns.  At 500 ns the signal has decayed somewhat less completely compared to the low density measurements.  All fluorescence curves are in plotted in arbitrary units (linear scale) and have been normalized with a peak of 1.
  }
  \label{fig:twodens}
\end{figure}

After the plasma is generated, plasma ions are excited using a cw laser beam tuned to the $4s \; ^2S_{1/2} \rightarrow 4p \; ^2P_{1/2}$ transition at 397 nm.  Fluorescence at this wavelength is collected using a lens, isolated using a band-pass optical interference filter, detected using a photomultiplier tube, and recorded using a fast digital oscilloscope.  While most of the fluorescence from the upper $^2P_{1/2}$ state is at 397 nm, there is also a 6\% branch to a dark,  metastable $3d \; ^2D_{3/2}$ state.  The 397 nm laser beam is focused to a Gaussian waist of $\sim 100\;\mu$m in the center of the plasma.  Typical fluorescence curves are shown in Fig. \ref{fig:twodens}.

Because the ions are all initially in the ground state, it takes time for them to start scattering photons from the probe laser beam.  In the fluorescence signal, this is seen as the rapid initial rise at time $t=0$.
As mentioned already, the ions initially accelerate due to the forces from other ions in the plasma.  This leads to disorder-induced heating.  This causes a rapid fall in the fluorescence signal after about 30 ns.

The plasma also expands because it is no longer confined by the MOT.  When the initial electron energy is not too low, this plasma expansion velocity is approximated fairly well by the equation
\begin{equation}
  v_{\rm exp}(r,t) = r\; \frac{2 k_B T_e}{m \sigma^2} \; t, \label{eqn:accel}
\end{equation}
where $r$ is the radial coordinate, $k_B$ is Boltzmann's constant, $T_e$ is the electron temperature, $m$ is the ion mass, and $t$ is time.  This equation is valid only for very early times in the plasma evolution, before the plasma has expanded, and before the electron temperature has changed due to expansion, evaporation, or recombination.  This expression also neglects the influence of correlations, velocity-changing collisions, or many-body effects.  A characteristic time for plasma expansion is when $v_{\rm exp}t = \sigma$.  Substituting the rms value of $r=\sqrt{3}\sigma$, we find that for $T_e = 57$ K, the characteristic time is approximately 2 $\mu$s.  As the plasma expands, the fluorescence continues to fall.  In our simulation, we can neglect the overall plasma expansion by concentrating on the first few hundred ns of data.

As can be seen in Fig. \ref{fig:twodens}, the fluorescence method clearly distinguishes between high and low density.  Disorder-induced heating occurs more rapidly in high density plasmas than in low density plasmas.  The fluorescence signal shows a weak oscillation around 100 ns in high density plasmas that is absent in low density plasmas.  Finally, at 500 ns, the fluorescence signal does not decay as completely for high density plasmas as it does for low density plasmas.  For a fixed density, the fluorescence signal also changes as the initial electro temperature is reduced.  If this signal can be properly interpreted, details about electron shielding, three-body recombination, many-body effects, and particle correlations can be determined at very early times in the plasma evolution.

\section{Simulation}

We simulate the plasma ion fluorescence signal by integrating the optical Bloch equations for a collection of ions in a cell.  In the simulation, ions interact via the Yukawa potential,
\begin{equation}
  V(r)= \frac{q_0^2}{4\pi\epsilon_0} \;\frac{e^{-r/\lambda_D}}{r},
  \label{eqn:yukawa}
\end{equation}
where $q_0$ is the fundamental charge, $\lambda_D = \sqrt{k_B T_e \epsilon_0/n q_0^2}$ is the Debye length, $k_B$ is Boltzmann's constant, $T_e$ is the electron temperature, $n$ is the density, and $\epsilon_0$ is the permittivity of free space.

Plasma ions are randomly distributed over a cubic cell. The cell dimension is $L \approx \sigma/10$.  For the Gaussian density profile of Eq. \ref{eqn:dens}, the density is constant across this cell size.  We differentiate Eq. \ref{eqn:yukawa} to find the $x,y,$ and $z$ components of the force on each ion due to the presence of all of the other (screened) ions in the cell.  We use a fourth-order Runge Kutta stepper to move the ions in time.

At each time step, we also solve the optical Bloch equations for each ion in the cell.  The ions are approximated as two-level atoms.  The Hamiltonian for an ion is written in terms of the detuning $\omega$ and the Rabi frequency $\Omega$ as
\begin{equation}
  H = \hbar \omega \sigma_{-} \sigma_{+} + \frac{\hbar \Omega}{2}\left( \sigma_{-} + \sigma_{+}\right) ,
\end{equation}
\noindent where the the raising and lowering matrices are
\begin{equation}
  \begin{array}{cc}
    \sigma_{+} = \left( \begin{array}{cc}
      0 & 1 \\ 0 & 0
    \end{array}\right)
  &
    \sigma_{-} = \left( \begin{array}{cc}
      0 & 0 \\ 1 & 0
    \end{array}\right)
  \end{array}.
\end{equation}
The equation for the density matrix is
\begin{eqnarray}
  \dot{\rho} = & -& i\omega \left( \sigma_+ \sigma_- \rho - \rho \sigma_+ \sigma_- \right) \nonumber \\
    & - & i \frac{\Omega}{2} \left[ \left( \sigma_+ + \sigma_- \right)\rho - \rho \left( \sigma_+ + \sigma_- \right) \right] \label{eqn:dmat} \\
    & + & \frac{\gamma}{2} \left(  2\sigma_- \rho \sigma_+ - \sigma_+ \sigma_- \rho - \rho \sigma_+\sigma_-  \right). \nonumber
\end{eqnarray}
where $\gamma$ is the total decay rate of the $4p \; ^2P_{1/2}$ state.  This equation of motion is somewhat more compactly written using the Pauli spin matrices, with $\sigma_x = \sigma_{+} + \sigma_{-}$, $\sigma_y = -i\sigma_{+} + i\sigma_{-}$, and $\sigma_z$ the $2\times 2$ identity matrix:
\begin{eqnarray}
  \frac{d\left< \sigma_z \right>}{dt} &=& \Omega \left< \sigma_y \right> - \gamma \left( 1 + \left<\sigma_z \right>\right) \\
  \frac{d\left< \sigma_y \right>}{dt} &=& \omega \left< \sigma_x \right> - \Omega \left< \sigma_z \right> - \frac{\gamma}{2}\left< \sigma_y \right> \\
  \frac{d\left< \sigma_x \right>}{dt} &=& -\omega \left< \sigma_y \right> - \frac{\gamma}{2}\left< \sigma_x \right>.
\end{eqnarray}
The fraction of atoms in the excited state is given by $\sigma_z$.  For example, if there is no 397 nm laser, then $\Omega=0$ and the equation for $\sigma_z$ is
\begin{equation}
  \left<\sigma_z\right>(t) = \left[ 1+ \left< \sigma_z\right>(0) \right]e^{-\gamma t} - 1.
\end{equation}
With the 397 nm laser present, $\Omega \neq 0$ and the fluorescence signal depends on time as
\begin{equation}
  f(t) = \frac{1}{2}\left[ 1 + \left< \sigma_z(t)\right> \right].
\end{equation}

Because the Bloch equations depend on the detuning $\omega$ of the laser beam from the 397 nm resonance transition, we need to consider two frequency shifts.  The first is the motion of the ions in the cell.  This is given by the first-order Doppler shift,
\begin{equation}
  \Delta \omega = \frac{2\pi\nu}{c}\;v,
\end{equation}
where $\nu$ is atomic transition frequency in the rest frame of the atom, $c$ is the speed of light, and $v$ is the component of the atomic velocity along the direction of the laser beam propagation.

The second frequency shift comes from the radial acceleration of the plasma (Eq. \ref{eqn:accel}).  For early enough times in the plasma evolution, the ions accelerate and begin to move, but the overall density profile is unchanged.  Therefore we can simulate the frequency shift due to plasma expansion by placing our cell at different plasma radii, calculating the mean velocity according to Eq. \ref{eqn:accel}, and adding this directed velocity to all of the ion velocities in the cell.

For each time step, we calculate the position and velocity of each ion.  We use wrapped boundary conditions to maintain a constant number of ions in the cell.  We use the velocity (including the effect of plasma expansion) to determine the ion's frequency shift $\omega$ and use a second-order Runge-Kutta routine to calculate its $\sigma_z$.  The $\sigma_z$ is then averaged over the ions in the cell to simulate the fluorescence signal.  The simulation is performed for many different initial ion densities. The ion density is chosen randomly from a Gaussian distribution so that the different density regions are appropriately weighted.

Care must be taken to properly simulate the fluorescence decay into the optically dark $3d$ state.  At each time step the total fluorescence rate from the  $^2P_{1/2}$ level is calculated and multiplied by the branching ratio to get the decay rate into the dark $3d \; ^2D_{3/2}$ level.  This decay rate is multiplied by the time step $dt$ to get the probability that the ion has made a transition to this level.  As is traditionally done in Monte Carlo simulations, this probability is compared to a random number with a flat distribution between 0 and 1.  If the random number is less than the probability, then the simulated ion makes the transition to the dark state and no longer fluoresces.  Decay into the dark metastable state reduces the fluorescence signal, particularly at longer times.

\section{Comparison with experiment}

\begin{figure}
  \includegraphics[width=3.35in]{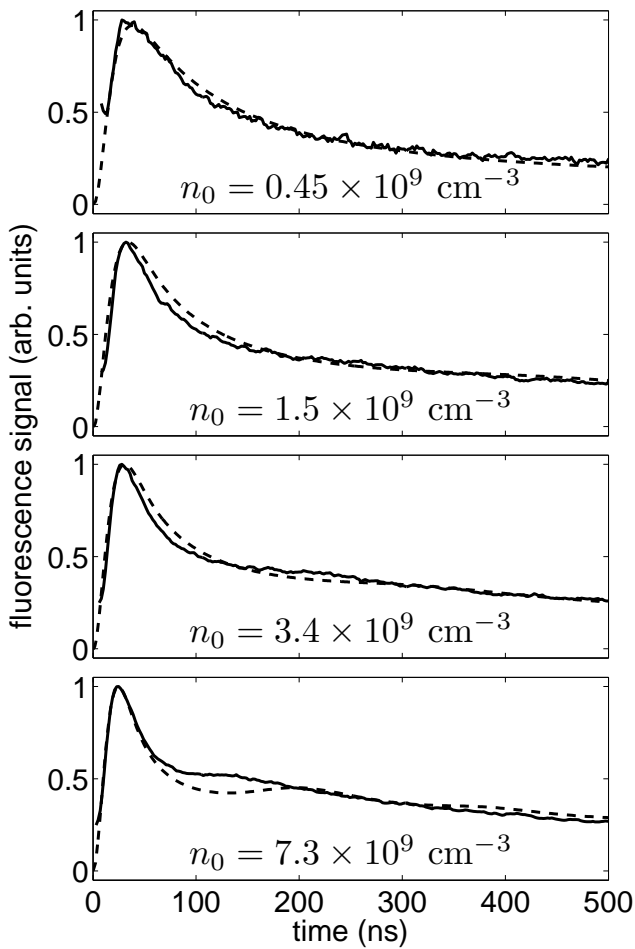}
  \caption{Comparison of experiment (solid line) with simulation (dashed line) for a range of initial plasma densities.  Both the simulation and experiment show initial rotation of the Bloch vector into the excited state, disorder-induced heating, and plasma expansion.  As the density increases, the simulation shows the oscillation in the rms velocity distribution occurring later in time than in the experiment.  This may be due to the onset of many-body interactions.  For all of these measurements, the initial electron temperature is $T_e=57$ K.}
  \label{fig:sigcomp}
\end{figure}

The simulation is compared with the experimental fluorescence signal in Fig. \ref{fig:sigcomp}.  The three phases of the signal are clearly visible. The first is the rapid rise of the fluorescence.  This corresponds to the initial rotation of the $z$-component of the Bloch vector into the excited state.  The second is an almost equally rapid decline in fluorescence after 20 ns.  This is due to disorder-induced heating -- the acceleration of the ions as they move to minimize their electrical potential energy.  The third is the slower signal decay at longer times.  This is due to the overall plasma expansion and to shelving into the $3d$ state.

The agreement between the simulation and experiment is quite good.  It suggests that the essential physics of the ion acceleration and plasma expansion are captured in this relatively simple approach.  In particular, the initial Bloch vector rotation and disorder-induced heating are well-simulated.

There is an apparent discrepancy in the ion temperature oscillation at high density.  As has been noted previously, the ions are all initially at rest.  They begin to move in their local electrical potential energy well.  After a time approximately equal to the inverse of the plasma frequency, the ions on average have moved to the other side of their local well and come to rest as they ``collide'' with their nearest neighbor.   This motion -- initially at rest, simultaneous acceleration, coming to rest again after one plasma period -- produces an oscillation in the rms ion velocity distribution.  Because nearest neighbors are randomly located in the initial plasma, the local potential is not parabolic and the collision time is not exactly the same for the ions.  After a short time, the coherence in the rms ion velocity oscillation is lost and the fluorescence curve smooths out.

This discrepancy may be due to collisions or optical pumping in the plasma.  Velocity changing collisions and additional shifts from the MOT magnetic fields, and many-body interactions may also influence the fluorescence signal.  These possibilities are currently under study.

\section{Conclusion}

We present measurements and simulations of laser-induced fluorescence of ions in an ultracold neutral plasma.  In the simulation, the ions interact via the Yukawa potential.  The ion position and velocity is calculated as a function of time, including the influence of ions in the cell as well as the overall plasma expansion.  At each time step we solve the optical Bloch equations and track the excited-state population to simulate the fluorescence signal.  We find good agreement between simulation and experiment.

Fluorescence measurements can be used to probe the early-time ion dynamics over a wide range of initial plasma densities.  The ability to accurately calculate fluorescence properties of ultracold plasmas makes it possible to quantitatively interpret laser-induced fluorescence signals from ultracold plasmas.  Exploring the influence of an external magnetic field, probe laser detuning, varying optical depth in the plasma, displacing the probe laser beam from the center of the plasma, and other important effects can be done as straightforward extensions of our approach.

Future work will focus on the agreement at higher densities in the plasma.  This regime is important to understand because it is where many-body effects are likely to appear.  Similarly, work at lower initial electron temperature is also interesting for the same reason.  At the very lowest initial electron temperature or at the highest plasma densities, the energy scale in the simulation is determined only by the disorder-induced heating, both for electrons and ions.  Comparing the Yukawa simulation, the experiment, and a direct molecular dynamics calculation would provide good insight into the onset of correlations in this regime.

This work is supported in part by the Research Corporation, the National Science Foundation (Grant No. PHY-0601699) and the Chemical Sciences, Geosciences, and Biosciences Division of the Office of Basic Energy Sciences, U.S. Department of Energy.

\end{document}